\documentclass[pra,showpacs,showkeys,amsfonts,amsmath,twocolumn]{revtex4}
\usepackage{bm}
\usepackage{graphicx}
\RequirePackage{mathptm}

\numberwithin{equation}{section}
\newcommand{\si}[1]{\sigma_{#1}}

\newcommand{\W}[4]{\begin{cases}
#1 ,&#2\\
#3 ,&#4
\end{cases}}
\newcommand{\ro}{\rho}
\newcommand{\la}{\lambda}

\newcommand{\be}{\beta}
\newcommand{\g}{\gamma_{0}}

\newcommand{\Ga}{\Gamma}
\newcommand{\om}{\omega_{0}}

\newcommand{\I}{\mathbb I}
\newcommand{\ket}[1]{|{#1}\rangle}
\newcommand{\bra}[1]{\langle {#1} |}
\newcommand{\cH}{{\mathcal H}}
\newcommand{\C}{\mathbb C}

\newcommand{\fA}{\mathfrak A}

\newcommand{\tr}{\mathrm{tr}\,}
\newcommand{\ptr}[1]{\mathrm{tr}_{#1}}

\newcommand{\mr}[1]{\mathrm{#1}}
\newcommand{\DS}{\displaystyle}
\begin{document}
\title{Noise-induced finite-time disentanglement in two-atomic
system}
\author{Lech Jak{\'o}bczyk}
\affiliation{Institute of Theoretical Physics\\ University of
Wroc{\l}aw\\
Pl. M. Borna 9, 50-204 Wroc{\l}aw, Poland}
\author{A. Jamr{\'o}z}
\affiliation{Institute of Theoretical Physics\\ University of
Wroc{\l}aw\\
Pl. M. Borna 9, 50-204 Wroc{\l}aw, Poland}
\begin{abstract}
We discuss the influence of a noisy environment on entangled states
of two atoms and show that all such states disentangle in finite
time.
\end{abstract}
\maketitle
\section{Introduction}
Entanglement of quantum states is the most non-classical feature of
quantum systems. It shows up when the system consists of two (or
more) subsystems and the total state cannot be written as a product
state. This notion can be generalized to mixed states, and mixed
state is non-separable or entangled if a corresponding density
matrix cannot be expressed as convex combination of tensor products
of density matrices of subsystems \cite{Werner}. Pure entangled
states as superpositions of multiparticle states, are fragile with
respect to noise resulting from interaction with environment. So to
control the effects of noise, it is important to understand details
of the process of \textit{disentanglement} i.e. to analyse how
entanglement can be destroyed by this interaction.
\par
In the paper, we describe the model of two two-level atoms
interacting with thermal bath at formally infinite temperature. When
atoms are separated by large distance, we can assume that the atoms
are located inside two independent baths. In such a case time
evolution of two-atomic system is given, in the Markovian
approximation, by the ergodic dynamical semi-group i.e. a trace
preserving semi-group of completely positive operators which have
the maximally mixed state $\frac{ 1}{ 4}\,\I_{4}$ invariant. Density
matrix corresponding to the state of the system satisfies the master
equation which right hand side is given by the Lindblad generator of
the semi-group. Since two atoms are largely separated, the
dipol-dipol interaction and the photon exchanges between atoms are
negligible, so in our model the generator is parametrized only by
the dissipation rate $\Ga$.
\par
From the general properties of this kind of evolution it follows
that the entanglement as a function of time always decreases to
zero, and that dynamics needs only a finite time to disentangle any
initially entangled state. Thus for all times $t$ greater then some
\textit{time of disentanglement} $t_{\mr{d}}(\ro)$, the states
$\ro(t)$ are separable. This finite-time effect should be compared
with asymptotic noise-induced decoherence effects (see also
\cite{D,YuE}). We calculate time evolution of some class of initial
density matrices and obtain analytic expression for its
entanglement. We also find  formulas for $t_{\mr{d}}$ in the cases
of pure initial states and some  mixed initial states. In that
examples, the time of disentanglement is the increasing function of
the initial entanglement. Another interesting aspect of dissipative
evolution studied in the paper is connected with nonlocal properties
of quantum states. Nonlocality of quantum theory manifests by
violation of Bell inequalities, and in the case of two two-level
atoms can be quantified by some parameter ranging from $0$ for local
states to $1$ for states maximally violating some Bell inequality.
Our dynamics enables also to consider evolution of this parameter.
In particular, we show that the time after which all nonlocal
properties of quantum state is lost is much shorter then the time of
disentanglement.
\section{Time evolution in a noisy environment}
\subsection{Two-level atom in a noisy environment}
Time evolution of a density matrix of two-level atom  $A$ inside the
bath with finite  temperature $T$ can be described by the following
master equation
\begin{equation}
\begin{split}
\frac{\DS d\ro}{\DS dt}=&\frac{\DS 1}{\DS 2}\,\Ga_{\uparrow}\,\{
[\si{+},\ro\si{-}]+[\si{+}\ro,\si{-}]\}+\\
& \frac{\DS 1}{\DS 2}\, \Ga_{\downarrow}\,\{
[\si{-},\ro\si{+}]+[\si{-}\ro,\si{+}]\}
\end{split}
\end{equation}
where
$$
\si{\pm}=\frac{\DS 1}{\DS 2}\,(\si{1} \pm \,i \si{2})
$$
and we identify ground state $\ket{0}$ and excited state $\ket{1}$
of the atom with vectors $\left(\begin{array}{c}
 0\\
 1\\
\end{array}\right)$ and $ \left(\begin{array}{c}
1\\
0\\
\end{array}\right)$ in $\cH_{A}=\C^{2}$. Moreover,
\begin{equation}
\Ga_{\uparrow}=\g \, n(\om),\quad \Ga_{\downarrow}=\g\, (1+n(\om))
\end{equation}
where
$$
n(\om)=\frac{\DS 1}{\DS e^{\be \om} -1}, \quad \be=\frac{\DS 1}{\DS
T}
$$
$\g$ is a spontaneous emission rate of the atom and $\om$ is the
frequency of the transition $\ket{0}\to \ket{1}$. Since
$$
\frac{\DS \Ga_{\downarrow}}{\DS \Ga_{\uparrow}}=e^{\be \om}\to 1
$$
when $\be \to 0$, for very high temperature ($T\to \infty$) we can
assume that \cite{Y}
$$
\Ga_{\uparrow}=\Ga_{\downarrow}=\Ga
$$
In this case (II.1) reduces to
\begin{equation}
\frac{\DS d\ro}{\DS dt}=L_{\Ga}\ro=\Ga\,
(\si{+}\ro\si{-}+\si{-}\ro\si{+}-\ro)
\end{equation}
and $L_{\Ga}$ generates the semi-group
$$
T_{t}=e^{t L_{\Ga}}
$$
which is ergodic i.e. $\{ T_{t} \}_{t\geq 0}$ has a unique
asymptotic state which is maximally mixed state $\frac{\DS
\I_{2}}{\DS 2}$ in $\C^{2}$. So the relaxation process described by
(II.3) which models the evolution of two-level atom inside the bath
with very high (infinite) temperature, brings all initial states of
the atom into the state with maximal entropy. In the other words,
the semi-group generated by (II.3) describes \textit{ open quantum
system} (two level atom) interacting with \textit{noisy environment}
(a bath with very high temperature).
\subsection{Two independent two-level atoms}
In the case of two separated two-level atoms $A$ and $B$ located
inside independent baths with formally infinite temperatures, the
generator of the corresponding semi-group is given by the following
generalization of (II.1)
\begin{equation}
\begin{split}
\frac{\DS d\ro^{AB}}{\DS dt}=L_{\Ga}^{AB}\ro^{AB}=&\Ga\,
(\si{+}^{A}\ro^{AB}\si{-}^{A}+\si{-}^{A}\ro^{AB}\si{+}^{A} +\\
&\si{+}^{B}\ro^{AB}\si{-}^{B}+\si{-}^{B}\ro^{AB}\si{+}^{B}-2\ro^{AB})
\end{split}
\end{equation}
where $\ro^{AB}$ is the state of the compound system $AB$, described
by the Hilbert space $\cH_{AB}=\cH_{A}\otimes \cH_{B}=\C^{4}$ and
the algebra of observables
$$
\fA_{AB}=M_{2\times 2}(\C)\otimes M_{2\times 2}(\C)\simeq M_{4\times
4} (\C)
$$
and
$$
\si{\pm}^{A}=\si{\pm}\otimes \I,\quad \si{\pm}^{B}=\I\otimes
\si{\pm}
$$
The semi-group generated by $L_{\Ga}^{AB}$ is also ergodic with
unique asymptotic state $\frac{\DS \I_{4}}{\DS 4}$ in $\C^{4}$. To
simplify the discussion  of the evolution of two-atomic system, let
us introduce the basis of so called collective states in the Hilbert
space $\cH_{AB}=\C^{4}$ \cite{FT}. If
$$
f_{1}=\ket{1}\otimes\ket{1},\, f_{2}=\ket{1}\otimes\ket{0},\,
f_{3}=\ket{0}\otimes\ket{1},\, f_{4}=\ket{0}\otimes\ket{0}
$$
then this basis containing excited state, ground state  and
symmetric and antisymmetric combination of the product states, is
defined as follows
\begin{equation}
\ket{e}=f_{1},\, \ket{g}=f_{4},\, \ket{s}=\frac{\DS 1}{\DS
\sqrt{2}}\,(f_{2}+f_{3}),\, \ket{a}=\frac{\DS 1}{\DS
\sqrt{2}}\,(f_{2}-f_{3})
\end{equation}
From the master equation (II.4) it follows that the matrix elements
with respect to the basis $\ket{e},\, \ket{s},\, \ket{a},\, \ket{g}$
of the state $\ro$ satisfy
\begin{equation}
\begin{split}
\frac{d\ro_{aa}}{dt}=&-2\Ga\, \ro_{aa}+\Ga\,
(\ro_{ee}+\ro_{gg})\\[2mm]
\frac{d\ro_{ss}}{dt}=&-2\Ga\,
\ro_{ss}+\Ga\,(\ro_{ee}+\ro_{gg})\\[2mm]
\frac{d\ro_{gg}}{dt}=&-2\Ga\, \ro_{gg}+\Ga\,(\ro_{ss}+\ro_{aa})\\[2mm]
\frac{d\ro_{ee}}{dt}=&-2\Ga\, \ro_{ee}+\Ga\,
(\ro_{ss}+\ro_{aa})\\[2mm]
\frac{d\ro_{eg}}{dt}=&-2\Ga\, \ro_{eg}\\[2mm]
\frac{d\ro_{as}}{dt}=&-2\Ga\, \ro_{as}\\[2mm]
\frac{d\ro_{ae}}{dt}=&-2\Ga\,\ro_{ae}-\Ga\, \ro_{gs}\\[2mm]
\frac{d\ro_{ag}}{dt}=&-2\Ga\, \ro_{ag}-\Ga\, \ro_{ea}\\[2mm]
\frac{d\ro_{se}}{dt}=&-2\Ga\,\ro_{se}+\Ga\, \ro_{gs}\\[2mm]
\frac{d\ro_{sg}}{dt}=&-2\Ga\, \ro_{sg}+\Ga\, \ro_{es}
\end{split}
\end{equation}
Notice that if the initial state belongs to the class of density
matrices of the form
\begin{equation}
\ro=
\begin{pmatrix}
\ro_{ee}&0&0&\ro_{eg}\\
0&\ro_{ss}&\ro_{sa}&0\\
0&\ro_{as}&\ro_{aa}&0\\
\ro_{ge}&0&0&\ro_{gg}
\end{pmatrix}
\end{equation}
then $\ro(t)$ given by the solution of the equations (II.6) also
belongs to that class. In this case one finds that
\begin{equation}
\begin{split}
\ro_{aa}(t)=&\frac{1}{4}+\frac{1}{2}\,e^{-\,2\,\Ga\,t}\,(\ro_{aa}-\ro_{ss})
+ \frac{1}{2}\,e^{-\,4\,\Ga\,t}\,(\ro_{aa}+\ro_{ss}-\frac{1}{2}\,) \\
\ro_{ss}(t)=&\frac{1}{4}-\frac{1}{2}\,e^{-\,2\,\Ga\,t}\,(\ro_{aa}-\ro_{ss})
+ \frac{1}{2}\,e^{-\,4\,\Ga\,t}\,(\ro_{aa}+\ro_{ss}-\,\frac{1}{2}\,)\\
\ro_{ee}(t)=&\frac{1}{4}+\frac{1}{2}\,e^{-\,2\,\Ga\,t}\,(\ro_{ee}-\ro_{gg})
+ \frac{1}{2}\,e^{-\,4\,\Ga\,t}\,(\ro_{ee}+\ro_{gg}-\,\frac{1}{2}\,)\\
\ro_{gg}(t)=&\frac{1}{4}-\frac{1}{2}\,e^{-\,2\,\Ga\,t}\,(\ro_{ee}-\ro_{gg})
+ \frac{1}{2}\,e^{-\,4\,\Ga\,t}\,(\ro_{ee}+\ro_{gg}-\,\frac{1}{2}\,)\\
\ro_{eg}(t)=&e^{-\,2\,\Ga\,t}\,\ro_{eg}\\
\ro_{as}(t)=&e^{-\,2\,\Ga\,t}\,\ro_{as}
\end{split}
\end{equation}
\section{Entanglement in two-atomic systems}
\subsection{Measure of entanglement}
In the case when subsystems of the total system are described by
two-dimensional Hilbert spaces, the natural measure of the amount of
entanglement a given quantum state contains i.e. the entanglement of
formation \cite{Bennett}
\begin{equation}
E(\ro)=\min \, \sum\limits_{k}\la_{k}E(P_{k})
\end{equation}
where the minimum is taken over all possible decompositions
\begin{equation}
\ro=\sum\limits_{k}\la_{k}P_{k}
\end{equation}
and
\begin{equation}
E(P)=-\tr[ (\ptr{A}P)\,\log_{2}\, (\ptr{A}P)]
\end{equation}
can be analytically computed as a function of another quantity
$C(\ro)$ called \textit{concurrence}, which also can be taken as a
measure of entanglement \cite{HW, W}. $C(\ro)$ is defined as follows
\begin{equation} C(\ro)=\max\;
(\,0, 2p_{\mathrm{max}}(\widehat{\ro})-\tr \widehat{\ro}\,)
\end{equation}
where $p_{\mathrm{max}}(\widehat{\ro})$ denotes the maximal
eigenvalue of $\widehat{\ro}$ and
\begin{equation}
\widehat{\ro}=(\ro^{1/2}\tilde{\ro}\ro^{1/2})^{1/2}
\end{equation}
with
\begin{equation}
\tilde{\ro}=(\si{2}\otimes \si{2})\,\overline{\ro}\,(\si{2}\otimes
\si{2})
\end{equation}
The value of the number $C(\ro)$ varies from $0$ for separable
states, to $1$ for maximally entangled pure states. Consider now the
class (II.7) of density matrices. With respect to the canonical
basis $f_{1},\, f_{2},\, f_{3},\, f_{4}$, the matrices (II.7) have
also the same form i.e.
$$
\ro=\begin{pmatrix} \ro_{11}&0&0&\ro_{14}\\
0&\ro_{22}&\ro_{23}&0\\
0&\ro_{32}&\ro_{33}&0\\
\ro_{41}&0&0&\ro_{44}
\end{pmatrix}
$$
One can check that for this class
\begin{equation}
C(\ro)=\max\, \{0,C_{1},C_{2} \}
\end{equation}
where
\begin{equation}
C_{1}=2\,(|\ro_{14}| - \sqrt{\ro_{22}\ro_{33}}\,),\quad C_{2}=2\,
(|\ro_{23}|- \sqrt{\ro_{11}\ro_{44}}\,)
\end{equation}
or
\begin{equation}
\begin{split}
C_{1}=&2\,|\ro_{eg}|-\sqrt{(\ro_{aa}+\ro_{ss})^{2}-(\ro_{as}+\ro_{sa})^{2}}\\[2mm]
C_{2}=&\sqrt{(\ro_{ss}-\ro_{aa})^{2}-(\ro_{as}-\ro_{sa})^{2}}-2\,
\sqrt{\ro_{ee}\ro_{gg}}
\end{split}
\end{equation}
when we use the matrix elements with respect to the collective
basis. In the special case when $\ro_{14}=0$, $C_{1}$ cannot be
positive, so
$$
C=\max \,\{0,C_{2}\}
$$
Similarly, when $\ro_{23}=0$, then
$$
C=\max \, \{0,C_{1} \}
$$
\subsection{Evolution of entanglement}
Suppose that the initial state of the two-atomic system belongs to
the class (II.7). Since the evolution given by the master equation
(II.4) leaves this class invariant, to compute entanglement at time
$t$ we can use formulas (III.9) and (II.8). So we have
$$
C(\ro(t))=\max\, \{0,C_{1}(t),\,C_{2}(t)\}
$$
where
\begin{widetext}
\begin{equation}
C_{1}(t)=2e^{-2\Ga t}|\ro_{eg}|- \sqrt{\left[\,e^{-4\Ga
t}\,\,(\,\ro_{aa}\,+\,\ro_{ss}-
\frac{1}{2}\,)+\frac{1}{2}\,\right]^{2}-e^{-4\Ga
t}(\,\ro_{as}+\ro_{sa}\,)^{2}}
\end{equation}
and
\begin{equation}
\begin{split}
C_{2}(t)=&e^{-2\Ga
t}\sqrt{(\ro_{ss}-\ro_{aa})^{2}-(\ro_{as}-\ro_{sa})^{2}}\,-\\[2mm]
&\frac{1}{2}\sqrt{1+e^{-8\Ga
t}(-1+2\ro_{ee}+2\ro_{gg})^{2}+4e^{-4\Ga t}\,\left[\ro_{ee}+
\ro_{gg}-\frac{1}{2}-(\ro_{ee}-\rho_{gg})^{2}\,\right]}
\end{split}
\end{equation}
\end{widetext}
As an example, consider the following initial state
\begin{equation}
\ro=\begin{pmatrix} 0&0&0&0\\
0&\ro_{22}&\ro_{23}&0\\
0&\ro_{32}&\ro_{33}&0\\
0&0&0&0
\end{pmatrix}
\end{equation}
The initial entanglement is equal to
$$
C(\ro)=2\,|\ro_{23}|=\sqrt{(\ro_{ss}-\ro_{aa})^{2}-(\ro_{as}-\ro_{sa})^{2}}
$$
so
\begin{equation}
C(\ro(t))=\max\, \{ 0, e^{-2\Ga t}C(\ro)-\frac{1}{2}\,(1-e^{-4\Ga
t})\}
\end{equation}
From (III.13) we see that there is the time $t_{\mr{d}}(\ro)$  after
which $C(\ro(t))$ becomes equal to $0$. We may call
$t_{\mr{d}}(\ro)$ \textit{the time of disentanglement} of a given
initial state $\ro$. For initial state (III.12)
\begin{equation}
t_{\mr{d}}(\ro)=\frac{1}{2\Ga}\ln\,
\left[C(\ro)+\sqrt{1+C(\ro)^{2}}\,\right]
\end{equation}
As we show in the next section, the appearance of finite time of
disentanglement for all initially entangled states, is the
characteristic feature of the dynamics governed by the master
equation (II.4).
\section{Time of disentanglement}
The evolution of states given by the semi-group  generated by
$L_{\Ga}^{AB}$ has the following important properties:
\\[2mm]
\textbf{(i)} it is local i.e. if $\ro$ is separable, then $\ro(t)$
is also separable for all $t\geq 0$,\\[2mm]
\textbf{(ii)} every initial state $\ro$ evolves to a maximally
mixed state $\frac{1}{4}\I_{4}$.\\[2mm]
Since maximally mixed state is separable and there is a
neighbourhood of this state which contains only separable states,
for an arbitrary state $\ro$ the set
\begin{equation}
S_{\ro}=\{ t\in [0,\infty)\,:\, \ro(t)\quad\text{is separable}\, \}
\end{equation}
is always non-empty. Moreover, if
\begin{equation}
E_{\ro}=\{ t\in [0,\infty)\,:\, \ro(t)\quad\text{is entangled}\, \}
\end{equation}
then $S_{\ro}$ and $E_{\ro}$ are disjoint and $S_{\ro}\cup
E_{\ro}=[0,\infty)$. Notice that for every $t_{1}\in E_{\ro}$ and
every $t_{2}\in S_{\ro}$ we have $t_{1}<t_{2}$, so $S_{\ro}$ is
bounded from below. Now we can define \textit{the time of
disentanglement} $t_{\mr{d}}(\ro)$ of a state $\ro$ of two-atomic
system as follows \cite{BJO}:
\begin{equation}
t_{\mr{d}}(\ro)=\inf\, S_{\ro}
\end{equation}
Since the set of separable states is compact, $S_{\ro}$ is closed
and $t_{\mr{d}}(\ro)\in S_{\ro}$. Therefore $t_{\mr{d}}(\ro)$ may be
also defined as the smallest time for which $\ro(t)$ is separable.
From the above discussion it is clear that for every initial state
of the system, there exist \textit{finite} time of disentanglement
(which may be equal to $0$ for separable initial states). Consider
now some explicite examples.\\[4mm]
\textbf{1.} Pure initial states.\\[2mm]
Since all pure entangled states $\Psi$ with concurrence $C(\Psi)=c$
are locally equivalent to the state given by the vector (see e.g.
\cite{L})
\begin{equation}
\Phi=\frac{1}{\sqrt{2}}\,\left(\,\sqrt{1+\sqrt{1-c^{2}}},\,0,\,0,\,
\sqrt{1-\sqrt{1-c^{2}}}\,\right)
\end{equation}
and our dynamics is local, it is enough to consider (IV.4) as
initial state. The corresponding density matrix
\begin{equation}
P_{\Phi}=\frac{1}{2}\,\begin{pmatrix} 1+\sqrt{1-c^{2}}&0&0&c\\[1mm]
0&0&0&0\\[1mm]
0&0&0&0\\[1mm]
c&0&0&1-\sqrt{1-c^{2}}
\end{pmatrix}
\end{equation}
evolves into
\begin{widetext}
\begin{equation}
P_{\Phi}(t)=\frac{1}{4}\,\begin{pmatrix} 1+e^{-4\Ga
t}+2\sqrt{1-c^{2}} e^{-2\Ga t}&0&0& 2c \, e^{-2\Ga t}\\[1mm]
0&1-e^{-4\Ga t}&0&0\\[1mm]
0&0&1-e^{-4\Ga t}&0\\[1mm]
2c\,e^{-2\Ga t}&0&0&1+e^{-4\Ga t}-2\sqrt{1-c^{2}}\,e^{-2\Ga t}
\end{pmatrix}
\end{equation}
\end{widetext}
One can easily check that
\begin{equation}
t_{\mr{d}}(P_{\Phi})=\frac{1}{2\Ga}\ln\,
\left[c+\sqrt{1+c^{2}}\,\right]
\end{equation}
This finite time of disentanglement should be contrasted with
infinite time needed for decoherence process (see  \cite{D,YuE} for
discussion of different models). For pure initial states, we can
introduce \textit{a decoherence rate} $\lambda (P)$ which tells us
how fast given pure initial state becomes mixed during the evolution
\cite{BJO}. If as a measure of mixedness we take the linear entropy
$$
S_{\mr{lin}}(\ro)=1-\tr \ro^{2}
$$
then
\begin{equation}
\lambda (P)=\frac{1}{2}\,\frac{d S_{\mr{lin}}(P(t))}{dt}\,\bigg
|_{t=0}
\end{equation}
In the case of state (IV.5) one obtains
\begin{equation}
\lambda (P_{\Phi})= 2\Ga
\end{equation}
We see also that off-diagonal elements of $P_{\Phi}(t)$ vanish
asymptotically with that rate.\\
Another characteristic time of our evolution is connected with so
called quantum non-locality i.e. the possibility of violating Bell
inequalities in quantum states \cite{Bell, CHSH}. It is known that
all pure entangled states violate Bell inequalities \cite{Gisin},
but this is not true for mixed states. In the case of two  two-level
systems, there is an effective criterion for violating  Bell
inequalities in mixed states \cite{HHH,H}. For any density matrix
$\ro$, take the real $3\times 3$ matrix
$$
T_{\ro}=\left( t_{nm} \right),\quad t_{nm}=\tr
(\ro\,\si{n}\otimes\si{m})\, n,m=1,2,3
$$
Define also real symmetric matrix
\begin{equation}
U_{\ro}=T_{\ro}^{T}\,T_{\ro}
\end{equation}
with eigenvalues $u_{1},\, u_{2},\, u_{3}$. Then $\ro$ violates some
Bell inequality if and only if \cite{HHH}
\begin{equation}
m(\ro)>1
\end{equation}
where
\begin{equation}
m(\ro)=\max_{j<k}\; (u_{j}+u_{k})
\end{equation}
As a measure of nonlocality we may introduce a function
$$
n(\ro)=\max\, \{0,\, m(\ro)-1 \}
$$
Initial state (IV.5) violates Bell inequality since
\begin{equation}
m(P_{\Phi})=1+c^{2}
\end{equation}
On the other hand,
\begin{equation}
m(P_{\Phi}(t))=e^{-8\Ga t}+c^{2}e^{-4\Ga t}
\end{equation}
 decreases to $0$, so there exists the time $t_{\mr{loc}}$ after
which (IV.14) is smaller then $1$ and nonlocality of initial state
is lost. We see that this \textit{locality time} for the state
(IV.5) equals
\begin{equation}
t_{\mr{loc}}(P_{\Phi})=\frac{1}{4\Ga}\, \ln\,\left[\frac{\DS
c^{2}+\sqrt{4+c^{4}}}{\DS 2}\, \right]
\end{equation}
and is always smaller then the time of disentanglement
$t_{\mr{d}}(P_{\Phi})$ (see FIG. 1). \\[2mm]
\begin{figure}
\centering
\includegraphics[width=80mm]{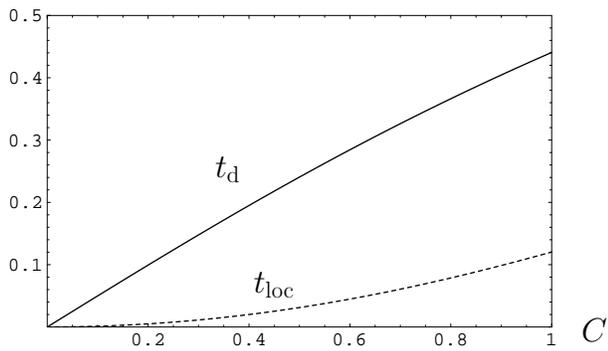}
\caption{$t_{\mr{d}}$ and $t_{\mr{loc}}$ in units of $[1/\Ga]$ as
functions of concurrence $C$, for pure initial states}
\end{figure}
We see that the evolution of pure initial states has the following
remarkable properties:\\[2mm]
\textbf{(a)} in the time interval $[0,t_{\mr{loc}})$ the states
(IV.6) are entangled and violate Bell inequalities,\\[2mm]
\textbf{(b)} for $t\in [t_{\mr{loc}},t_{\mr{d}}),\; P_{\Phi}(t)$ are
still entangled but do not violate any Bell inequality,\\[2mm]
\textbf{(c)} for all $t\geq t_{\mr{d}}$ the states (IV.6) are
separable, although decoherence process takes infinite time.\\[4mm]
\textbf{2.} Some mixed initial states.\\[2mm]
\textbf{(i)} Consider the class of Werner states \cite{BBP}
\begin{equation}
W_{\pm}=(1-p)\,\frac{\DS \I_{4}}{\DS
4}+p\,\ket{\Psi_{\pm}}\bra{\Psi_{\pm}}
\end{equation}
where
$$
\Psi_{\pm}=\frac{\DS 1}{\DS
\sqrt{2}}\,\left[\,\ket{0}\otimes\ket{0}\pm
\ket{1}\otimes\ket{1}\,\right]
$$
are maximally entangled pure states. It is known that $W_{\pm}$ are
entangled for $p>1/3$ and $C(W_{\pm})=\frac{\DS 3p-1}{\DS 2}$.
During the time evolution $W_{\pm}$ become
\begin{widetext}
\begin{equation}
W_{\pm}(t)=\frac{1}{4}\,\begin{pmatrix} 1+p\,e^{-4\Ga t}&0&0&\pm
2p\,e^{-2\Ga t}\\[1mm]
0&1-p\, e^{-4\Ga t}&0&0\\[1mm]
0&0&1-p\, e^{-4\Ga t}&0\\[1mm]
\pm 2p\, e^{-2\Ga t}&0&0&1+p\, e^{-4\Ga t}
\end{pmatrix}
\end{equation}
\end{widetext}
and
\begin{equation}
C(W_{\pm}(t))=\max\, \left\{ 0,p\, \left(\,e^{-2\Ga t}+\frac{1}{2}\,
e^{-4\Ga t}\,\right) -\frac{1}{2}\,\right\}
\end{equation}
One finds that
\begin{equation}
t_{\mr{d}}(W_{\pm})=\frac{1}{2\Ga}\, \ln\,\left[
p+\sqrt{p(1+p)}\,\right],\quad p>\frac{1}{3}
\end{equation}
On the other hand, not all Werner states which are entangled,
violate Bell inequalities. Nonlocal properties have only those
states $W_{\pm}$ with $p>\frac{\DS 1}{\DS \sqrt{2}}$ \cite{HHH}.
This nonlocality is lost when $t>t_{\mr{loc}}(W_{\pm})$
\begin{equation}
t_{\mr{loc}}(W_{\pm})=\frac{1}{4\Ga}\,\ln 2 p^{2},\quad
p>\frac{1}{\sqrt{2}}
\end{equation}
But even in the interval $\frac{\DS 1}{\DS \sqrt{2}} <p\leq 1$, this
time is much smaller then time of disentanglement (FIG. 2)\\[2mm]
\begin{figure}
\centering
\includegraphics[width=80mm]{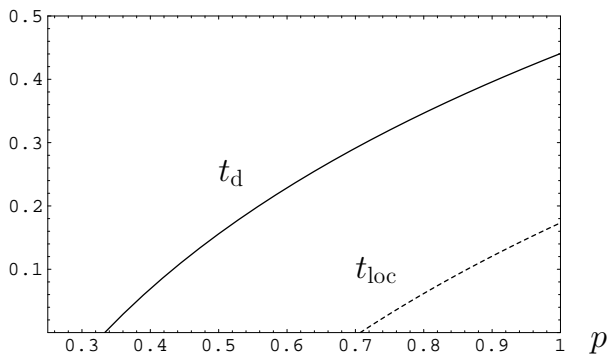}
\caption{$t_{\mr{d}}$ and $t_{\mr{loc}}$ in units of $[1/\Ga]$ as
functions of $p$ for Werner initial states}
\end{figure}
\begin{figure}
\centering
\includegraphics[width=80mm]{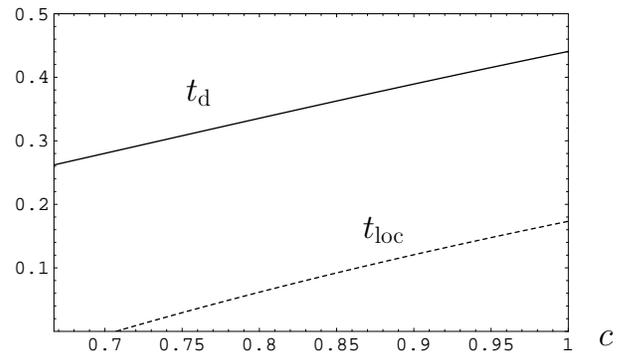}
\caption{$t_{\mr{d}}$ and $t_{\mr{loc}}$ in units of $[1/\Ga]$ as
functions of $c$ for maximally entangled mixed initial states}
\end{figure}
\textbf{(ii)} Similar computation can be done for the class of
maximally entangled mixed states \cite{M}, which have maximal value
of entanglement for a given degree of inpurity measured by linear
entropy
\begin{equation}
\ro_{\mr{MEMS}}=\begin{pmatrix} g(c)&0&0&c/2\\[1mm]
0&1-2g(c)&0&0\\[1mm]
0&0&0&0\\[1mm]
c/2&0&0&g(c)
\end{pmatrix}
\end{equation}
where
$$
g(c)=\W{1/3}{c\in [0,2/3]}{c/2}{c\in [2/3,1]}
$$
Direct calculations show that
\begin{widetext}
\begin{equation}
t_{\mr{d}}(\ro_{\mr{MEMS}})=\W{\frac{\DS 1}{\DS 4\Ga}\,
\ln\left[\,\frac{5}{9}
+2c^{2}+\frac{1}{16}\sqrt{(36c^{2}+10)^{2}-36}\,\right]}{c\in
[0,2/3]} {\frac{\DS 1}{\DS 4\Ga}\, \ln\left[
\,1-2c+4c^{2}+2\sqrt{2}c\sqrt{1-2c+2c^{2}}\,\right]}{c\in [2/3,1]}
\end{equation}
\end{widetext}
One can also check that $\ro_{\mr{MEMS}}$ violates Bell inequality
when $c>\frac{\DS 1}{\DS \sqrt{2}}$ and for such values of initial
entanglement, the locality time equals
\begin{equation}
t_{\mr{loc}}(\ro_{\mr{MEMS}})=\frac{1}{4\Ga}\, \ln\, 2c^{2},\quad
c>\frac{1}{\sqrt{2}}
\end{equation}
As in the previous cases, this time is always smaller then the time
of disentanglement (FIG. 3).\\[2mm]
We have shown that interaction of two-atomic system with a noisy
environment, modeled by the master equation (II.4) leads to the
disentanglement of initially entangled states in finite time. This
time of disentanglement is (at least in examples considered above)
the  increasing function of initial entanglement - more entangled is
the initial state, the longer period of time is needed to
disentangle it. On the other hand, if the initial entangled state
can violate Bell inequalities, the period of time in which it still
have nonlocal properties, is much shorter then the duration of the
process of disentanglement.

\end{document}